\documentclass[runningheads]{llncs}
\usepackage{todonotes}
\usepackage{graphicx}
\usepackage{multirow}
\usepackage{array}
\usepackage{cite}
\newcolumntype{P}[1]{>{\centering\arraybackslash}p{#1}}

\begin{document}

\title{Personality is Revealed During Weekends: Towards Data Minimisation for Smartphone Based Personality Classification} 

\titlerunning{Personality is Revealed During Weekends}

\author{Mohammed Khwaja\inst{1,2} \and
Aleksandar Matic\inst{1}}
\authorrunning{M. Khwaja \& A. Matic}
\institute{Telefonica Alpha, Barcelona, Spain\\
\email{name.surname@telefonica.com}\\
\and
Imperial College, London, UK}

\maketitle              

\begin{abstract}
Previous literature has explored automatic personality modelling using smartphone data for its potential to personalise mobile services. Although passive modelling of personality removes the burden of completing lengthy questionnaires, the fact that such models typically require a few weeks or months of personal data can negatively impact user's engagement. In this study, we explore the feasibility of reducing the duration of data collection in the context of personality classification. We found that only one or two weekends can suffice for achieving state-of-the-art accuracy between 66\% and 71\% for classifying the five personality traits. These results provide lessons for practicing "data minimisation" -- a key principle of privacy laws.  

\end{abstract}

\section{Introduction}

Personality reflects individual differences in behaviours, emotions and cognition~\cite{deyoung2009personality}. Psychologists showed that personality traits capture stable individual characteristics that explain and predict behavioural patterns~\cite{corr2009cambridge}. Interestingly, personality traits can also predict patterns of technology use, such as behaviours in social media~\cite{catal2017cross, ferwerda2018you}, blogs~\cite{minamikawa2011blog}, games~\cite{yee2011introverted}, phone use~\cite{chittaranjan2011s, staiano2012friends, monsted2018phone} and even how users choose app permission settings \cite{raber2017towards}. Therefore, personality is considered to be relevant to a number of computing areas, among which Human Computer Interaction (HCI) can particularly benefit from understanding users' personality, by making informed decisions about their needs and preferences. Consideration of personality was  shown to be highly beneficial for personalising recommender systems~\cite{ferwerda2016personality}, gamification elements~\cite{jia2016personality}, online educational applications~\cite{lee2017personalizing}, persuasive health games~\cite{orji2017towards} and other kinds of technologies. Previous work also demonstrated how personality influences adoption of new technologies~\cite{xu2016understanding} as well as users' satisfaction~\cite{oliveira2013influence}. 

Assessing personality typically relies on standardised questionnaires where individuals rate their typical behaviours with Likert scales. When it comes to user modelling, app designers typically avoid using this method as completing questionnaires can be cumbersome for users and can consequently drive them away from the app. For this reason, automatic  prediction of personality has attracted the attention of many scholars and practitioners who relied on data collected from Twitter~\cite{catal2017cross}, Instagram~\cite{ferwerda2018you}, blogs~\cite{minamikawa2011blog}, and smartphone use~\cite{chittaranjan2011s, staiano2012friends, monsted2018phone}. Most of these approaches relied on collecting data from several weeks~\cite{staiano2012friends}, months~\cite{oliveira2013influence} and even years~\cite{monsted2018phone, de2013predicting}, in order to accurately infer personality. 
In practice, however, collecting such large amounts of personal data is not always trivial. Firstly, data minimisation represents a fundamental principle of privacy both in the EU (under the General Data Protection Regulation - GDPR \footnote[1]{http://www.privacy-regulation.eu/en/article-5-principles-relating-to-processing-of-personal-data-GDPR.htm}) and in the US \cite{tene2012big}, which obliges organisations to collect only minimal amount of personal data for the intended purpose. Collecting large amounts of personal data was also shown to be strongly associated with low user engagement due to privacy concerns~\cite{staddon2012privacy}. Secondly, systems that rely on user data typically suffer from the "cold-start" problem~\cite{abel2010interweaving} -- in the case of personality prediction, requiring data collection of several weeks or months before enabling personalised services may be fatal for the user's engagement. These reasons underline the importance of understanding how to minimise data collection (or the data that is retained in the system) while at the same time reducing time needed to develop user models. This is what we explored using smartphone based personality classification.

In this study, we analysed if and to which extent the accuracy of personality inference will be affected when reducing the data collection to a few days (in contrast to weeks or months as in previous studies) and specifically to weekend days. The rationale for this study stems from the assumption that people exhibit more natural behaviours during weekends when they have more control over their time, than during working days. Zuzanek et al.~\cite{zuzanek2002life} argued that people engage in activities of their preference more frequently during weekends than weekdays, whereas Ryan et al.~\cite{ryan2010weekends} showed that mood is significantly better during weekends. To this end, the present study relies on 142 behavioural features extracted from two-week smartphone data collected from 166 participants to predict their Big Five personality traits. The main contributions of this paper are: 
\begin{itemize}
    \item A comparison between personality inferring machine learning models that rely on smartphone data collected during weekends versus weekdays.
    \item Takeaways for reducing duration of data collection (to one day, one weekend, and two weekends) for developing personality models.
\end{itemize}

\section{Background}
Extant literature explored personality inference approaches relying on various data from social network logs to keystroke patterns, and audio and video data. Considering the topic of this paper, we will provide an overview of the most important literature that relied on smartphone data to detect personality traits. A comprehensive review of personality modelling using various digital cues can be found in~\cite{vinciarelli2014survey}. 

Pioneering work in exploring phone data for personality prediction used call and message logs. Oliveira et al.~\cite{de2011towards} investigated structural characteristics of contact networks modelled through 6 months of call logs from 39 users, which resulted in promising preliminary results. Staiano et al.~\cite{staiano2012friends} extracted social network structures from 2 months of call logs and Bluetooth scans of 53 subjects and obtained binary classification accuracy between 65\% and 80\% for predicting the five traits. Chittaranjan et al.~\cite{chittaranjan2011s} and~\cite{chittaranjan2013mining} used 8 months of phone data (calls, messages, Bluetooth, and applications) of 83 and 117 subjects in two trials to predict personality; F-measures for the binary classification task was between 40\% and 80\%. Using call logs and location data of 69 participants, Montjoye et al.~\cite{de2013predicting} extracted psychology-informed indicators to predict personality between 29\% and 56\% better than random, relying on 12 months of data. Recent work by Monsted et al.~\cite{monsted2018phone} used 24 months of data from 636 university students to predict Extraversion. The authors used features from social activities extracted from calls, SMS, online networks, and physical proximity extracted from Bluetooth and GPS. Another recent research by Wang et al.~\cite{wang2018sensing} used mobile sensing data of 646 students from the University of Texas over 14 days to regress personality traits. This work used behavioural features like social interaction, movement, daily activity etc., from sensors including sound, activity, location and call logs to achieve Mean Average Error (MAE) between 0.39-0.61.

Past research provided a solid foundation of using smartphone data to infer personality traits, relying on datasets collected over several weeks and months to a few years. Yet, it remains unclear if data collection can be reduced in time while still achieving a comparable accuracy to the models developed using more longitudinal data. This would mitigate the cold-start problem and help service providers to enable data minimisation principles of privacy laws, while not sacrificing the quality of services. We believe that our work provides a contribution on that front. 

\section{Methodology}

\begin{table}
\centering
  \caption{Data Categories}
  \label{tab:data}
  \begin{tabular}{P{2.3cm}|P{7.5cm}|P{1.7cm}}
    \hline
    Category &  Description & Num of Daily Features\\
    \hline
    Light &  Provides the intensity of light in lux & 5\\
    Noise & Provides the level of noise in dB & 15\\
    Battery & Provides battery level, charging state & 2\\
    Accelerometer & Provides 3D acceleration during activity & 12\\
    Call & Provides duration, number and state of calls & 9\\
    Unlock & Provides screen on/off, phone lock state & 9\\
    Pedometer & Provides step count during an activity & 5\\
    Location & Provides GPS coordinates (Latitude and longitude) & 13\\
  \hline
\end{tabular}
\end{table}

For this work, we used data from 1) smartphone sensors (microphone, light, accelerometer, pedometer, location), 2) usage logs (phone unlocks, screen on/off, battery level and charging, calls), collected using an Android app - summarised in Table~\ref{tab:data}. The data sampling was optimised for a low battery consumption which resulted in no complaints from users about the battery consumption. Phone unlock events, screen on/off, battery charging logs and calls were captured for every event. Data from the microphone, pedometer, location and light sensors was collected every 15 minutes, while data from the accelerometer was sampled when it was detected that a person was moving. 

Participants were recruited through a specialised agency from February to August 2018. They were asked to install and keep the app active for 3 weeks, which was followed by completing a set of onboarding questionnaires that included demographics (gender, age, socioeconomic status, etc.) and the 50 item Big Five personality inventory~\cite{goldberg2006international}. Following the GDPR, participants were presented with details about the purpose of the study and the data collected, and were enrolled in the study only upon providing their consent. They also had the flexibility to decide which sensor information they would like to be recorded, which resulted in 69\% of participants providing partial data only. On successful completion, each participant received a monetary incentive of 40 EUR.
\vspace{-0.5ex}
\subsection{Participants}
\vspace{-0.8ex}
From over 1000 potential participants who were selected in this study trial, 545 participants from five countries successfully completed the study. However, due to missing sensor data the number of participants used for this analysis dropped to N=166 (Spain N=69, Peru N=25, Colombia N=21, Chile N=24 and the United Kingdom N=27). The gender ratio (female:male) for the eligible participants was roughly 1:2 and the age groups of the participants ranged between 18-25 (N=30), 26-34 (N=118) and 35-44 (N=18). Within each country, the gender ratio and age range ratio was roughly the same, as well as personality distributions. Importantly, distributions of the five personality scores with and without drop-outs did not significantly differ i.e. participants who dropped-out did not differ in personality from the rest of the sample. 

\subsection{Feature Extraction}
\vspace{-0.8ex}
Using the collected data, we first extracted a set of daily features that describe typical patterns of user behaviour and contexts during a day (e.g. mean level of light and noise during the morning or evening, distance travelled per day, radius of gyration, etc.) - similar to the previous literature \cite{vinciarelli2014survey, wang2018sensing}. Overall, 70 daily features were created from the categories described in Table~\ref{tab:data}, and each day was tagged as a weekday or a weekend day. Table~\ref{tab:data} also shows the number of daily features obtained from each category.

Using the tagged days, the data was then clustered into weekdays and weekend days. For each of the time periods, we also calculated the Routine Index, as defined in~\cite{canzian2015trajectories}. As participants typically finished participation during the third week of the study, we rarely collected the data from all three weeks at an individual level, and therefore we sub-sampled two weeks of data. We randomly selected four weekdays from the sub-sampled set, in order to use the same number of weekdays and weekend days when comparing the corresponding model accuracy. We aggregated the data during weekends and weekdays per participant, and extracted features by using descriptive statistics (mean and standard deviation) to describe typical behaviour during weekdays and during weekends. In this manner, we obtained 142 features for weekdays and 142 features for weekends.

\subsection{Model}

We approach personality inference as a machine learning classification problem. We split participants into two classes -- above and below the median value of the Big Five scores (Table~\ref{tab:stats}) -- for each of the five traits. This approach yields two balanced classes for developing each of the five classification models, which was commonly applied in personality detection literature~\cite{chittaranjan2011s, chittaranjan2013mining, vinciarelli2014survey, staiano2012friends}. 

Initially, we tested several classifiers, including Support Vector Machine, Bayes Naive Classifier and Nearest Neighbour, and we chose Random Forest as it outperformed the other methods. Random Forest has already been used for classification of personality traits in ~\cite{chittaranjan2011s, staiano2012friends} - it is a technique that typically does not require an extensive parameter tuning and feature selection. However, due to the number of features (142) in our case, feature selection brought performance improvements. We performed Recursive Feature Elimination in each step of the leave-one-out train-test method, that we used for the classifier accuracy assessment. In this way, the classifier was sequentially trained with the data from all but one user, tested with the data from the "left-out" user, and this process was repeated for all the users. As the performance metrics, we report the accuracy (Acc) of the classifier and the Cohen's Kappa ($\kappa$) value. The $\kappa$ value represents the improvement over the random classification. As we used the median value to create two classes of users, random classification by assigning 1 value to all users produces Acc $\approx$ 50\% and $\kappa \approx 0$. 

\section{Results}

\subsection{Questionnaire Analysis}

The Big Five personality dimensions include Extraversion, Agreeableness, Conscientiousness, Neuroticism and Openness that are obtained from the 50 item International Personality Item Pool~\cite{goldberg2006international}. The questionnaire asks users to rate their behaviours from 1-5 on a Likert scale, and each of the five traits is assessed through 10 questions with the aggregated score ranging from 10 to 50. The statistics of the scores for the Big Five traits are summarised in Table~\ref{tab:stats}, and are comparable to past literature~\cite{staiano2012friends}. The scores also showed a good internal reliability, with Cronbach's alpha $>$ 0.7 for all traits, also being in line with previous literature~\cite{gow2005goldberg}.

\begin{table}
\centering
  \caption{Statistics for the Big Five personality Scores}
  \label{tab:stats}
  \begin{tabular}{P{1.8cm}|P{1.8cm}|P{1.8cm}|P{1.8cm}|P{1.8cm}|P{1.8cm}}
    \hline
    Statistic & Extraversion & Agree. & Consc. & Neuroticism & Openness\\
    \hline
    Mean & 30.01 & 39.50 & 34.17 & 29.34 & 36.81\\
    Std. dev. & 7.42 & 5.56 & 5.55 & 7.83 & 5.01\\
    Median & 31.0 & 40.0 & 34.0 & 30.0 & 37.0\\
    Max & 48.0 & 50.0 & 50.0 & 48.0 & 50.0\\
    Min & 10.0 & 14.0 & 18.0 & 10.0 & 19.0\\
  \hline
\end{tabular}
\end{table}

\subsection{Personality Trait Inference}

Table~\ref{results} presents the accuracy of the personality classification models  - note that we removed the results with Acc $< 65\%$ or $\kappa < 0.3$ (denoted as '-' in the table). Although lower accuracy results have been reported in previous work, we set the threshold of 65\% for  classification accuracy as sufficient, based on \cite{park2018simpler}. For comparison with the models that rely on reduced datasets, we first developed a 'reference' model by using features computed using the full data set -- 2 weeks of smartphone data collected during both weekdays and weekends. The reference model was able to accurately classify between 68\% and 73\% of users for Openness, Agreeableness, Extraversion, Conscientiousness and Neuroticism, in ascending order of performance, with $\kappa$ ranging from 0.34-0.46. Our methodology, and moreover the results, are highly consistent with state-of-the-art work in personality classification~\cite{chittaranjan2013mining, chittaranjan2011s, staiano2012friends}.
\vspace{-1ex}
\begin{table}
\centering
\caption{Results obtained from the prediction of personality traits}
\label{results}
\begin{tabular}{P{3.45cm}|P{0.75cm}|P{0.75cm}|P{0.75cm}|P{0.75cm}|P{0.75cm}|P{0.75cm}|P{0.75cm}|P{0.75cm}|P{0.75cm}|P{0.75cm}}
\hline
Model/Type of day used & \multicolumn{10}{c}{Personality trait}\\
    \cline{2-11}
     & \multicolumn{2}{c|}{Extra.} & \multicolumn{2}{c|}{Agree.} & \multicolumn{2}{c|}{Consc.} & \multicolumn{2}{c|}{Neur.} & \multicolumn{2}{c}{Open.}\\
     \cline{2-11}
     & Acc (\%) & $\kappa$ & Acc (\%) & $\kappa$ & Acc (\%) & $\kappa$ & Acc (\%) & $\kappa$ & Acc (\%) & $\kappa$\\
\hline
Full dataset (2 weeks) & 71 & 0.43 & 70 & 0.39 & 71 & 0.38 & 73 & 0.46 & 68 & 0.34\\\hline

Weekend (2 weeks) & 69 & 0.38 & 71 & 0.41 & 71 & 0.4 & 70 & 0.38 & 66 & 0.3\\

Weekday (2 weeks) & - & - & 67 & 0.34 & 70 & 0.38 & 68 & 0.36 & - & -\\\hline

Weekend (1 week) & 67 & 0.35 & 66 & 0.31 & 68 & 0.35 & 67 & 0.34 & - & -\\

Weekday (1 week) & - & - & - & - & 66 & 0.31 & - & - & - & -\\\hline

Saturday & - & - & - & - & 68 & 0.33 & 67 & 0.34 & - & -\\

Sunday & - & - & 67 & 0.33 & 66 & 0.28 & - & - & - & -\\

Random Weekday & - & - & - & - & - & - & 65 & 0.3 & - & - \\\hline
\end{tabular}
\end{table}
\vspace{-5ex}
\subsubsection{1. Weekend vs Weekdays Model} 

To compare the predictive power of weekends and weekdays, we developed two consistent classification models by using 142 features only from weekends and only from weekdays respectively. To allow for a fair comparison, we randomly selected an equal number (i.e. four) of weekdays for computing the features and repeated the classification 10 times to ensure that we covered all the combinations. We observed that the model based on behavioural features extracted during two weekends was able to classify all the five personality traits with accuracy comparable to the reference model that relied on 14 days of data -- with only 1-3\% difference. The reference model was built using features from both weekend and weekdays, however it appears to provide only a marginal improvement over the weekend model. The model that relied only on weekdays classified Agreeableness, Conscientiousness and Neuroticsm with 67\%, 70\% and 68\% respectively, while not reaching the threshold of 65\% in predicting Extraversion and Openness. The weekend model significantly outperformed the weekday model for Extraversion, Agreeableness and Openness (McNemar's test, $p<0.01$))

\subsubsection{2. One vs Two Weekends Model} 

To further attempt to reduce duration of smartphone data used for personality classification, we evaluated a classification model developed using the features extracted from one weekend only. The accuracy dropped in comparison to the two-weekends model and to the reference model by 2\% to 6\%, while not being able to detect Openness. However, the accuracy in detecting Extraversion, Agreeableness, Conscientiousness, and Neuroticism were above the threshold of 65\%, despite using only one weekend (i.e. the data from two weekend days). We also compared this model with a model that uses features computed from two randomly selected weekdays and we observed statistically significant differences for prediction of all five traits - Agreeableness (McNemar's test, $p<0.001$), Conscientiousness, Openness, Extraversion, Neuroticism (McNemar's test, $p<0.05$). This further indicates the value that weekend behaviours bring to the personality modelling in comparison to weekdays.   

\vspace{-2ex}
\subsubsection{3. One day Model}

Next we attempted to further reduce the dataset to one day. Given the results from one weekend data, we aimed to evaluate which of the two weekend days is more predictive of traits - Saturday or Sunday. We compare the two models by selecting a random Saturday and a random Sunday, and also a random weekday for comparison (as in the previous cases, we repeated this procedure 10 times). Interestingly, the Saturday model was able to predict Conscientiousness and Neuroticism, the Sunday model was able to predict Agreeableness, and Conscientiousness - with a moderately good accuracy above the threshold of 65\% . McNemar's test indicated that the models obtained from Saturday and Sunday were significantly different for Agreeableness and Neuroticism ($p<0.05$). A random weekday model was not capable of classifying 4 out of the 5 traits, reaching 65\% only for Neuroticism. We also attempted to classify the traits by specifically selecting a single day of the week (e.g. Monday). This produced inadequate results and are not reported here.

\vspace{-1ex}
\section{Discussion and Conclusion}
\vspace{-1ex}
Personality has been in the focus of HCI researchers for its importance in understanding user needs, preferences and satisfaction with technologies, as well as for building more personalised services. Our study provides evidence that (1) smartphone data collected during weekends has a stronger predictive power than  weekday data for inferring personality traits, (2) only 2-4 days of smartphone data can be enough for achieving state-of-the-art accuracy in personality classification. We believe that this work has two main implications -- takeaways for enabling data minimisation, that is one of the key principles in privacy as well as lessons for shortening the time period needed for delivering customised services based on personality. 

In multiple tests (Table~\ref{results}) we observed that the smartphone data collected during weekends was significantly more predictive for inferring personality traits. Interestingly, by using two weekends i.e. four weekend days, the accuracy was highly comparable with previous personality classification studies that relied on several weeks or months (in a few cases even years) of data. During weekends people typically have more control over their activities in comparison to working days, which was explored by social scientists but it is also not difficult to intuitively deduce some differences. This served as a rationale for our study in which the weekend behaviours turned out to be more informative of individuals' personality (note that the literature has not explored how personality is manifested during working versus non-working days). Our future research will explore if further improvements can be achieved by distinguishing working and non-working days at an individual level instead of weekend versus weekdays.  

In practical terms, using two weekends of data does not resolve the cold-start problem as the user would still need to wait for almost two weeks until the service models his/her personality and becomes more personalised. However, our findings suggest the possibility to reduce the data retained at the service side, as a user's engagement is frequently affected by privacy concerns related to the amount of collected data. Moreover, minimising the personal data required for delivering a service is a core component in privacy guidelines. Further research in this direction can also probe the sensor modalities that are more important for personality prediction over others. 

In the context of the cold-start problem, our results indicate that it is possible to detect 4 out of 5 traits with an accuracy of above 65\% by using one weekend, or 3 traits by using only one weekend day. In practice, if a user did not install a service just before the weekend, it would still require several days until the modelling has been completed, yet this process significantly reduces the time needed for the personality inference. 

We hope that our study will motivate further work on data minimisation approaches, not only because of privacy regulations but also to encourage applying principles of ethical computing. We also believe that our study will inspire psychologists to delve deeper into manifestation of personality during different days of the week. 

\vspace{-1.5ex}
\section*{Acknowledgements}
\vspace{-1ex}

This work has been supported by the European Union's Horizon 2020 research and innovation programme, under the Marie Sklodowska-Curie grant agreement no. 722561.

\bibliographystyle{splncs04}

\end{document}